\documentclass[prd,aps,letterpaper,floatfix,superscriptaddress,preprintnumbers,twocolumn,10pt]{revtex4-1}

\usepackage{dcolumn}
\usepackage{bm}
\usepackage[dvips]{graphicx}
\usepackage{epsfig}
\usepackage{amsmath, amsfonts, amssymb, slashed}

\newcommand{\be}{\begin{equation}}
\newcommand{\ee}{\end{equation}}
\newcommand{\bee}{\begin{equation*}}
\newcommand{\eee}{\end{equation*}}

\usepackage{color}

\newcommand{\bea}{\begin{eqnarray}}  
\newcommand{\eea}{\end{eqnarray}}

\newcommand{\bsvar}{\sigma_{b}^2}
\usepackage{multirow}
\usepackage{feynmp}
\usepackage{tikz}
\begin{document}

\title{ 
Exploring SMEFT in VH  with Machine
Learning}

\author{Felipe F. Freitas}
\affiliation{CAS  Key  Laboratory  of  Theoretical  Physics,  Institute  of  Theoretical  Physics,  Chinese  Academy of Sciences, Beijing 100190, China}
\author{Charanjit K. Khosa} 
\affiliation{Department of Physics and Astronomy, University of Sussex, Brighton BN1 9QH, UK}
\author{Ver\'onica Sanz} 
\affiliation{Department of Physics and Astronomy, University of Sussex, Brighton BN1 9QH, UK}
\email{charanjit.kaur@sussex.ac.uk}
\email{felipe in china}
\email{V.Sanz@sussex.ac.uk}
\date{\today}

\begin{abstract}
In this paper we study the use of Machine Learning techniques to exploit kinematic information in VH, the production of a Higgs in association with a massive vector boson. We parametrize the effect of new physics in terms of the SMEFT framework.  We find that the use of a shallow neural network allows us to dramatically increase the sensitivity to deviations in VH respect to previous estimates. We also discuss the relation between the usual measures of performance in Machine Learning, such as AUC or accuracy, with the more adept measure of Asimov significance. This relation is particularly relevant when parametrizing systematic uncertainties. Our results show the potential of incorporating Machine Learning techniques to the SMEFT studies using the current datasets.
\end{abstract}

\maketitle

\section{Introduction}

 The Particle Physics community holds high hopes of discoveries in the Large Hadron Collider (LHC), the machine colliding protons at the highest energies in an Earth laboratory.  Yet, after years of an intense effort searching for new phenomena, no clear evidence of new physics has been found. 
 
 To continue the search for new phenomena and improve the exploitation of the LHC data, we are shifting our focus from the low-hanging fruit, e.g. resonance searches, into more subtle (indirect) effects of new physics. A well-defined approach to develop the interpretation of data in terms of indirect probes is the framework of Effective Field Theories~\cite{EFT}, and in particular in the context of the Standard Model EFT (SMEFT)~\cite{SMEFT}.

 In a nutshell, the SMEFT is a consistent way of exploring new theories as deformations from the SM structures,  with a large number of possible SM deviations taken into account. 
 
As an example, in  the SMEFT approach the Higgs couplings to vector bosons $V=W,Z$ would be modified in the following way
 \bea
\eta_{\mu\nu} \, g m_V \Rightarrow \eta_{\mu\nu} \, g m_V - \frac{2 \, g \, c_{HW} }{m_W} \, p^{V}_{\mu} \, p^{V}_{\nu} + \ldots \label{eqcoupling}
\eea 
which in terms of Lagrangian terms would be equivalent to adding to the SM Lagrangian new terms suppressed by a scale of new physics
\bea
{\cal L}_{SM} \Rightarrow {\cal L}_{SM} + \frac{2 i g c_{HW}}{m_W^2} \, [D^\mu H^\dagger T_{2k} D^\nu H] \, W^k_{\mu\nu} + \ldots \label{eqL}
\eea 
where $H$ is the Higgs $SU(2)$ doublet and $W^k$ is the electroweak gauge boson triplet. 
 \begin{figure}[!h]
\begin{center}
\includegraphics[width=0.22\textwidth]{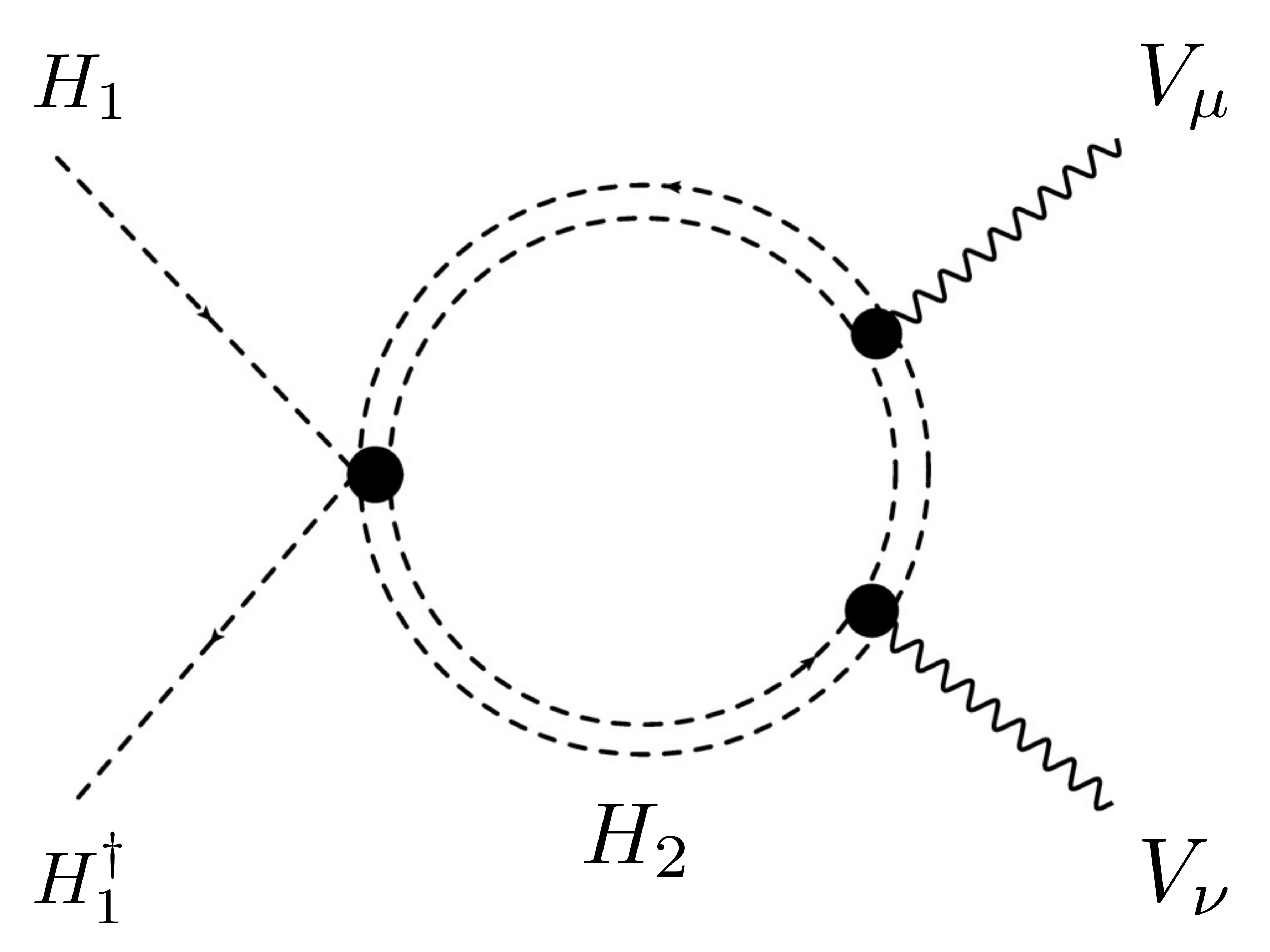} 
\caption{An example of how operators like $c_{HW}$ could arise from new theories. $H_{1,2}$ denotes the light (heavy) Higgs.}
\label{fig:oneloop}
\end{center}
\end{figure}

One could trace the ultimate origin of these deformations to many different types of new physics, just too heavy to be discovered directly at the LHC. For example, the deformation (aka Wilson coefficient) $c_{HW}$ could be the manifestation of a new set of scalar particles, such as in 2HDMs, too heavy or too complex to be seen in direct production, but still felt via virtual effects such as the one-loop contribution shown in Fig.~\ref{fig:oneloop}~\cite{Gorbahn}.

These new theories would then manifest themselves in the LHC environment as subtle deviations in physical observables, often in kinematic regions where the theoretical and experimental understanding is particularly poor. In contrast with a resonance search in a final state, SMEFT analyses forces us to deal with the LHC's inherently complex environment, where the understanding of extreme kinematic regions is required. 

In the context of the SMEFT effects in the Higgs sector, the LHC analyses have moved from the basic use of total cross-sections ($\kappa$ formalism~\cite{kappa}) to understand that pushing the boundaries of the SMEFT means using kinematic information~\cite{firstkinJohn}. Even that frontier is becoming a well-trodden path with the Run2 finished, and searches for new physics in SMEFT effects now moving towards identifying even subtler effects by looking at multidimensional information~\cite{sylvain} and combining as many channels as possible~\cite{johnnewest}. 

This state of affairs, the need to quickly identify subtle effects in multidimensional distributions of information, clearly calls for artificial intelligence methods. Particularly the use of data mining techniques in Machine Learning~\cite{MLreview}. The amount of information one single channel can provide is limited, though. Even in a complex final state such as Vector Boson Fusion (VBF) and all the multidimensional correlations one can think of in this channel, the amount of information quickly saturates~\cite{sylvain}, just a manifestation that the kinematics of the final state particles (input information) satisfies a number of constraints (energy-momentum conservation, behaviour of parton distribution functions, experimental selection cuts and resolution), limiting the usefulness of single channels. In VBF, we showed the inherent limitations in a Bayesian context~\cite{sylvain} and recently in Refs.~\cite{MLEFT}  the authors pioneered the use of Machine Learning to identify SMEFT effects in VBF, including data augmentation.

Despite its importance to understand the electroweak sector, the measurement of Higgs production in VBF is not a reality yet, hence studies are based on future prospects. On the other hand, the production of the Higgs in association with a massive vector boson, or VH, is already firmly stablished~\cite{Hbbatlas, HbbCMS}. As quality kinematic information in $WH$ and $ZH$,  better statistics and experimental understanding, will occur before VBF production is understood, we believe the approach of this paper would be the first step to push the boundaries of our understanding of SMEFT effects on Higgs LHC data, complemented later on with VBF information.  

In this paper, we will illustrate the use of Machine Learning techniques in VH by switching on a single SMEFT effect on $WH$ and $ZH$.  
In the past few months, we have witnessed an explosion works by the HEP community 
on the use of Machine Learning techniques, e.g. Refs.~\cite{MLHEPexample}, and the analyses have quickly become more and more sophisticated. Althoug in this paper we use state-of-the-art techniques, we expect our results in VH will be surpassed by other works in the near future. 

This paper is organised as follows. In Sec.~\ref{vh}, we describe the current status of the SMEFT analyses and the experimental understanding of the VH channel. In Sec.~\ref{kins}, we then move to describe the sort of kinematic information one could use in VH.  The Machine Learning analysis, in particular the use of a shallow neural network is described in Sec.~\ref{NNs} and Appendix~\ref{glossary}, where we provide a simple glossary of terms used in this paper. We present our results in Sec.~\ref{results}, and discuss possible new directions in Sec.~\ref{conclusions}.
\section{Current status:  limits on the SMEFT, and the VH at the LHC \label{vh}}
We are going to illustrate the techniques using a particular deformation, the operator in Eq.~\ref{eqL} with Wilson coefficient $d_{HW}$. It is currently constrained to values in the range~\cite{johnnewest} (individual constraint)
 \bea
 c_{HW} = 0.002 \pm 0.014 \ .
 \eea 
In this paper we will often illustrate points using a benchmark within the 2$\sigma$ region:
 \bea
c_{HW}=0.03 \ .
\eea
The limits on SMEFT operators were obtained by perfoming a global fit including kinematic information on VH \cite{johnhiggscomplete} and electroweak $WW$ production at LEP2 and LHC~\cite{johncompleteRun1} but only 40 fb$^{-1}$ of data, half of the total Run2 dataset. A more recent global analysis was done by the groups in Refs.~\cite{concha,tilmanglobalnew}, but their analysis did not substantially change the limit on $c_{HW}$. 
On the other hand, sensitivity studies of future colliders such as HL-LHC show that these limits will be pushed to a few times smaller than the current limit~\cite{HL-LHC}.

 On the experimental side, the ATLAS~\cite{Hbbatlas} and CMS~\cite{HbbCMS} collaborations have marked yet another milestone in their quest to understand electroweak symmetry breaking: the observation of the Higgs decaying into two b-quarks. This measurement has been done by combining a challenging set of channels collectively denoted by VH, which corresponds to the Higgs produced in association with a massive vector boson $V=$ $Z$ or $W^\pm$. The final  states are classified as 0L ($Z\to \nu \bar \nu$), 1L ($W \to \ell \nu$) and 2L ($Z\to \ell^+ \ell^-$). 
 The combination of all the channels can be summarised as the ratio of the observed cross-section by the SM expectation, $\mu_{VH}$. For example, the ATLAS measurement reads
\bea
\mu_{VH} \textrm{ (ATLAS) } = 1.01 + 0.12 \textrm{ (stat.)} ^{+0.6}_{0.15}\textrm{ (syst.)} 
\eea
which, given the dependence of $\mu$ with the parameter $c_{HW}$ naively would indicate a two-sigma exclusion $|c_{HW}|< $ 0.02.

\section{Kinematic information in VH\label{kins}}
Right after the discovery of the Higgs boson in the summer of 2012, the  VH channel was identified as an important source of information to search for anomalous behaviour of the Higgs. In particular, the distribution of transverse momentum of the vector boson, $p_T^V$ was identified as very sensitive to new physics, even to the point of reviving TeVatron searches which had failed to unveil the Higgs boson~\cite{firstkinJohn}. 

\begin{figure}[!h]%
    \begin{center}
   \includegraphics[scale = 0.23]{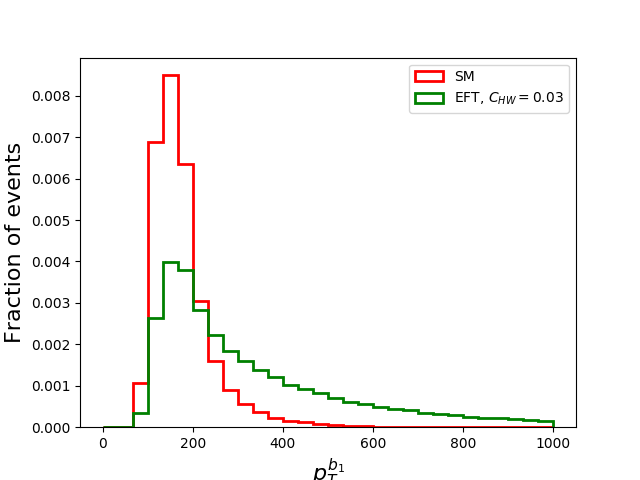}
  \includegraphics[scale = 0.23]{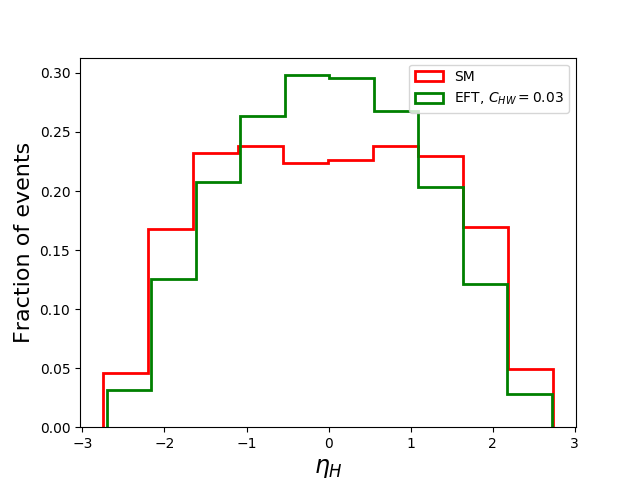}
  \includegraphics[scale = 0.23]{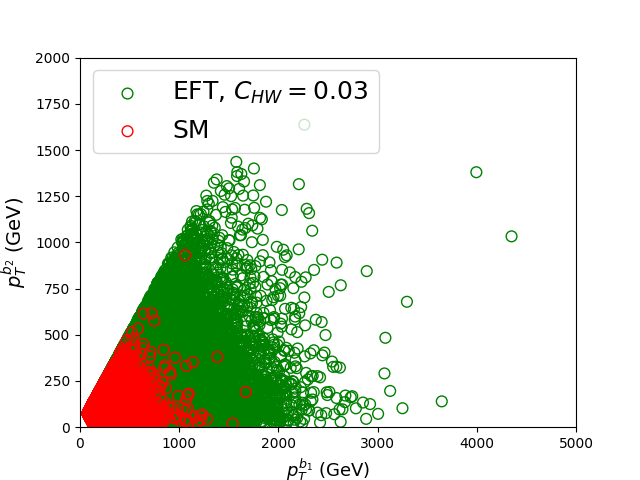}
  \includegraphics[scale = 0.23]{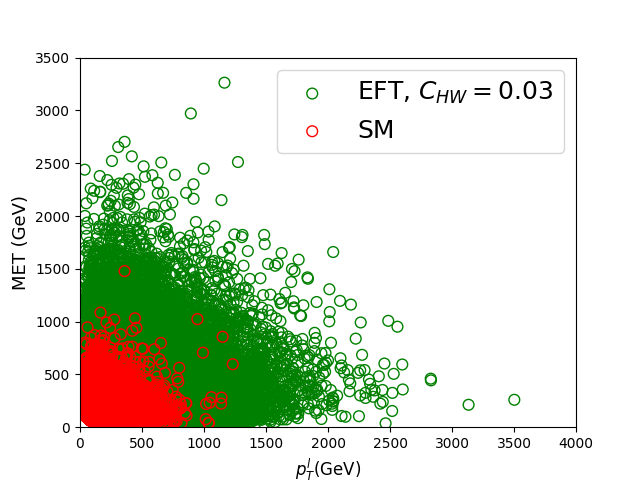}
   \includegraphics[scale = 0.23]{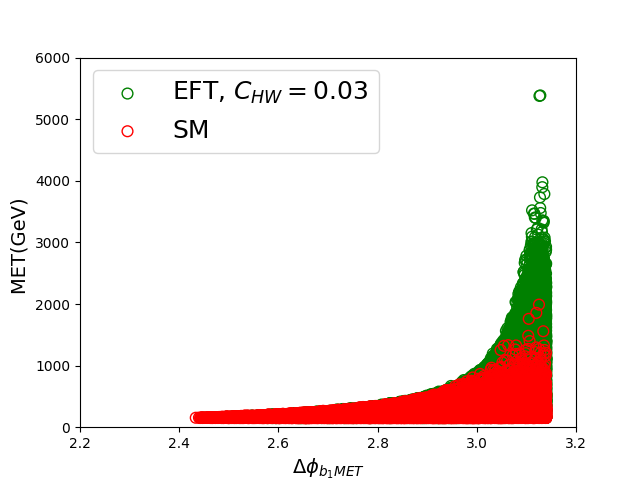}
  \includegraphics[scale = 0.23]{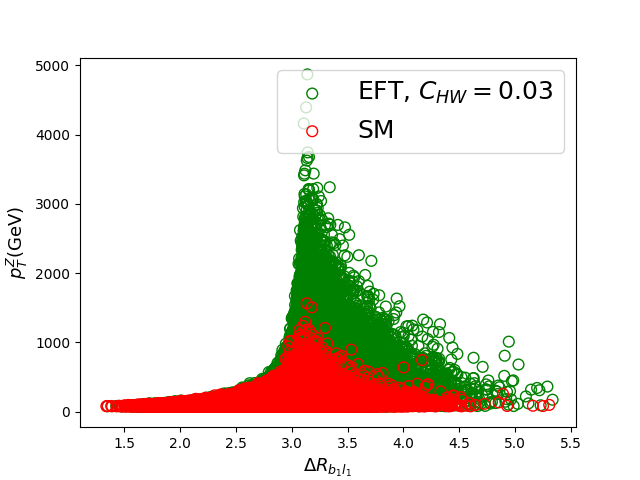}
  \includegraphics[scale = 0.23]{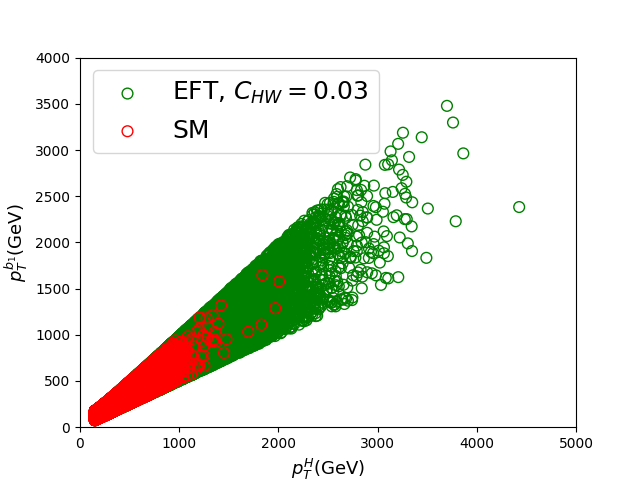}
  \includegraphics[scale = 0.23]{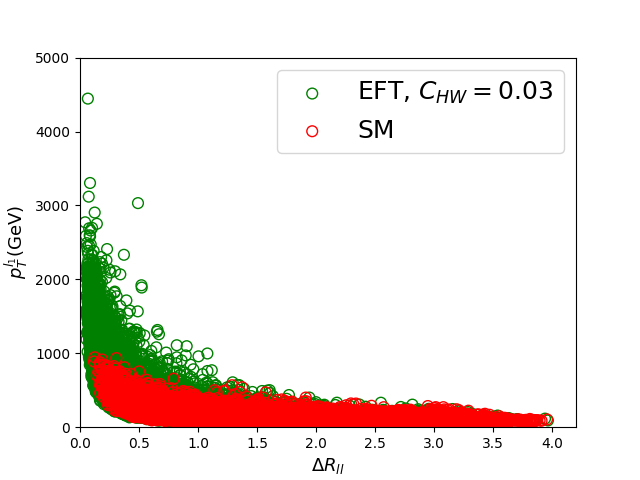}
  \end{center}
    \caption{Few illustrative 1D and 2D feature plots for inclusive 0L, 1L and 2L SM-Higgs production. Red dots correspond to background (SM Higgs production) and green dots to signal (Higgs SMEFT). Note the broader kinematic reach of the signal.}\label{fig:features}
\end{figure}

When Run1 LHC data started to place limits on the VH channel, the non-observation of deviations in the high-$p_T$ regions was also used to inform our global understanding of SMEFT theories~\cite{johnhiggscomplete, johncompleteRun1} and the CP properties of the Higgs~\cite{felipeCP}. Moreover, the understanding and classification of the $p_T^V$ distributions was crucial for Run2 collaborations to achieve a measurement~\cite{Hbbatlas,HbbCMS}.

As mentioned in the Introduction, the VH channel would seem qualitatively less interesting than the VBF channel, where the forward jets enrich the overall kinematic information. Nonetheless, VH with its slightly different three channels also offers interesting kinematic information, see in Fig.~\ref{fig:features} a few examples of distributions we will use later in our analysis.

Moreover, progress is made at stages and the experimental understanding of the VBF channel is nowhere close to VH. In VH,  huge SM reducible backgrounds, such as a $Z$ and heavy flavour production, had been studied and kept under control thanks to the tremendous ingenuity of the experimental collaborations. The recent observation of the Higgs sets then a new stage for the VH channel, where new physics can be searched and tensioned against SM-Higgs production. 

\section{Using a shallow neural network\label{NNs}}
In this section we will describe the methodology we developed to study the SMEFT in VH using Machine Learning, in particular a shallow neural network. To help the novice reader, in Appendix~\ref{glossary} we have collected a glossary of terms alongside brief explanation of their meaning.

To extract the maximum amount of information from the kinematic features, one needs to combine multidimensional information such as shown in Fig.~\ref{fig:features} in 0D, 1D, 2D and even higher dimensionalities. The objective is to {\it maximise} our ability to detect new phenomena, which in HEP means maximising the significance of an observation. Given a number of signal events $s$, where signal here represents the SM plus a deviation like $c_{HW}$, and a number of background events $b$, one can use the Asimov estimate of significance~\cite{CowanCranmer} as a measure to maximise, see Appendix~\ref{glossary}. Our problem then consists on building a function, inverse of the Asimov significance, and find its true minimum inside a complex parameter space by including information from a diverse set of observables.  

Similarly to the procedure described in Ref.~\cite{ellwood}, we use a shallow neural network (NN) built from one hidden layer with  number of neurons equal to the number of kinematic observables we consider. To set the best hyperparameters, instead of performing a brute force grid search as in ~\cite{ellwood}, we are making use of Evolutionary Algorithms in Python (DEAP)~\cite{DEAP}, in addition to the the Scikit-Learn library. As activation function, we found a rectifier function (max($x,0$)) to perform better than the typical sigmoid and other logistic regression options. For optimisation, we found Adam was best performing. Other minor adjustments were done to the batch size and the dropout options, see Appendix~\ref{glossary}. 

We then fed the algorithm with a large number of simulated events, both signal and background. The events had a number of characteristics, including $p_T$ of the objects (b-jets, leptons, missing energy) and combinations of different objects. The identity of the event (signal or background) was used by the algorithm as part of the training,  as we are dealing with a supervised machine learning problem. Before tackling minimisation of the Asimov loss function, we performed a pre-training set of runs for 5 epochs, along the lines suggested in Ref.~\cite{ellwood} using a steeper loss function. A longer run, with about 20-30 epochs was then done. 

The outcome of these runs was the ability to classify events as signal or background, and to assign a Asimov significance estimate to a particular choice of $c_{HW}$ coefficient (the strength of the deviation) and luminosity (the amount of available data). 
\begin{figure*}%
    \begin{center}
     \includegraphics[scale = 0.16]{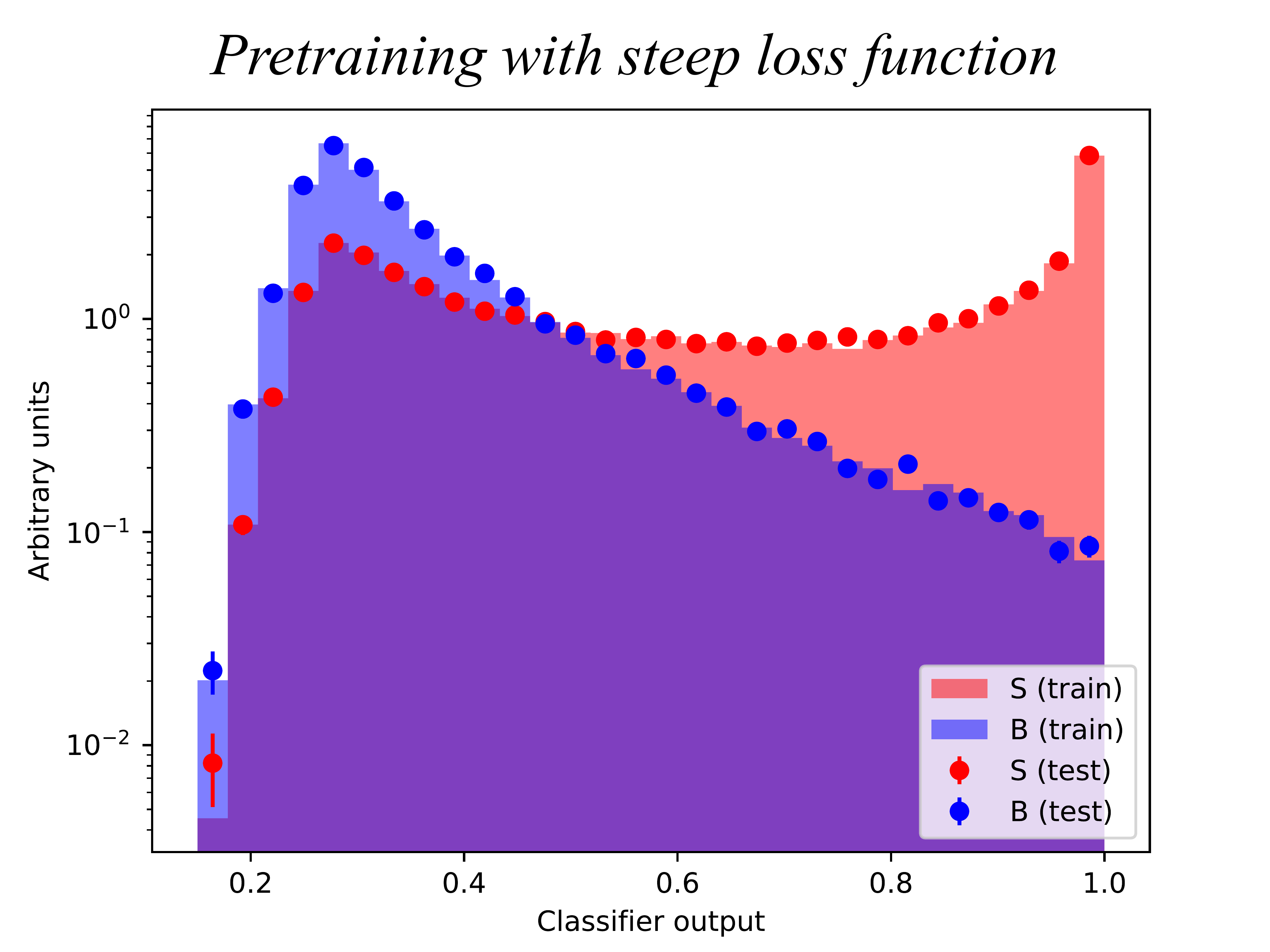} 
 \includegraphics[scale = 0.16]{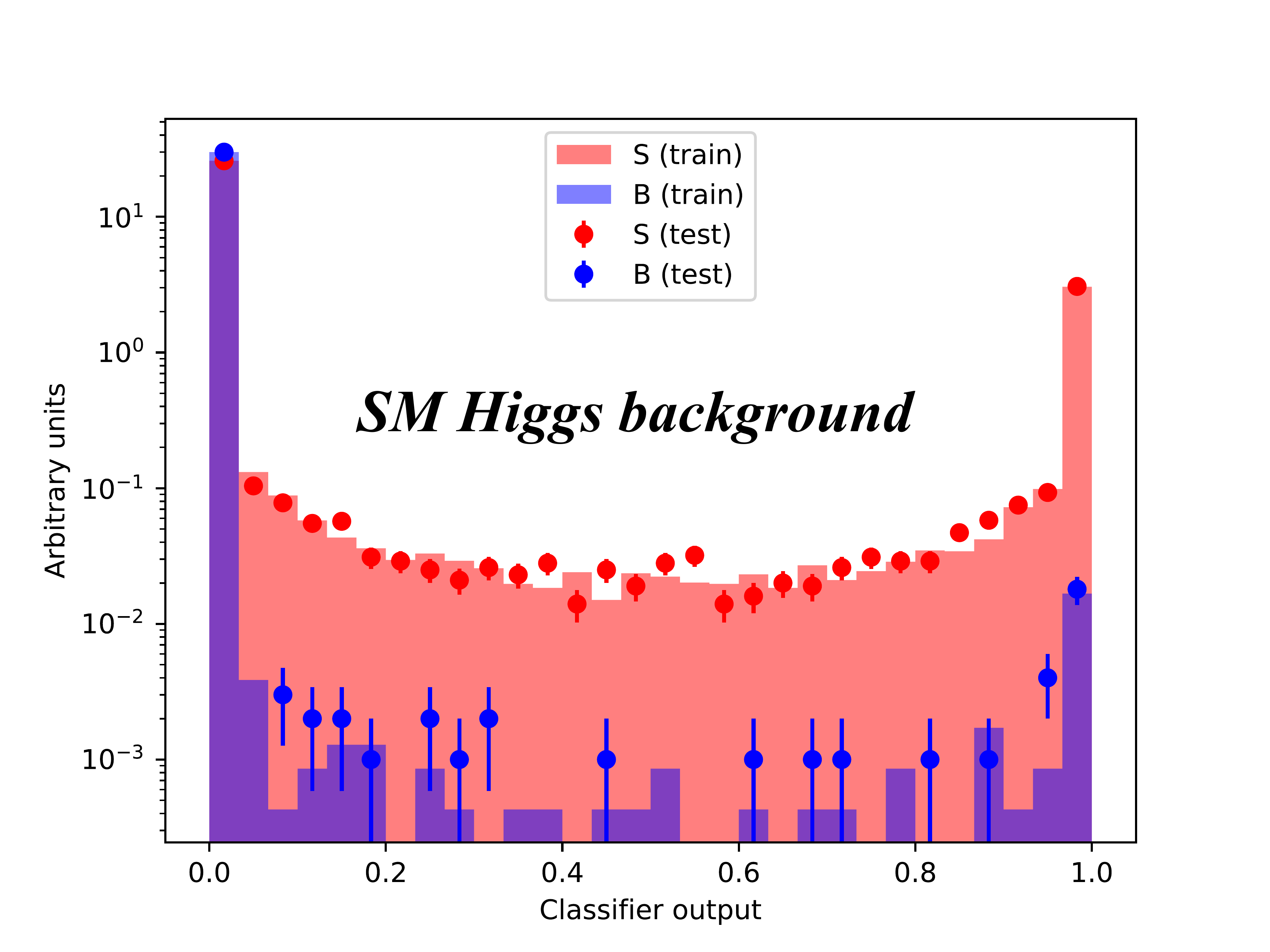} 
     \includegraphics[scale = 0.16]{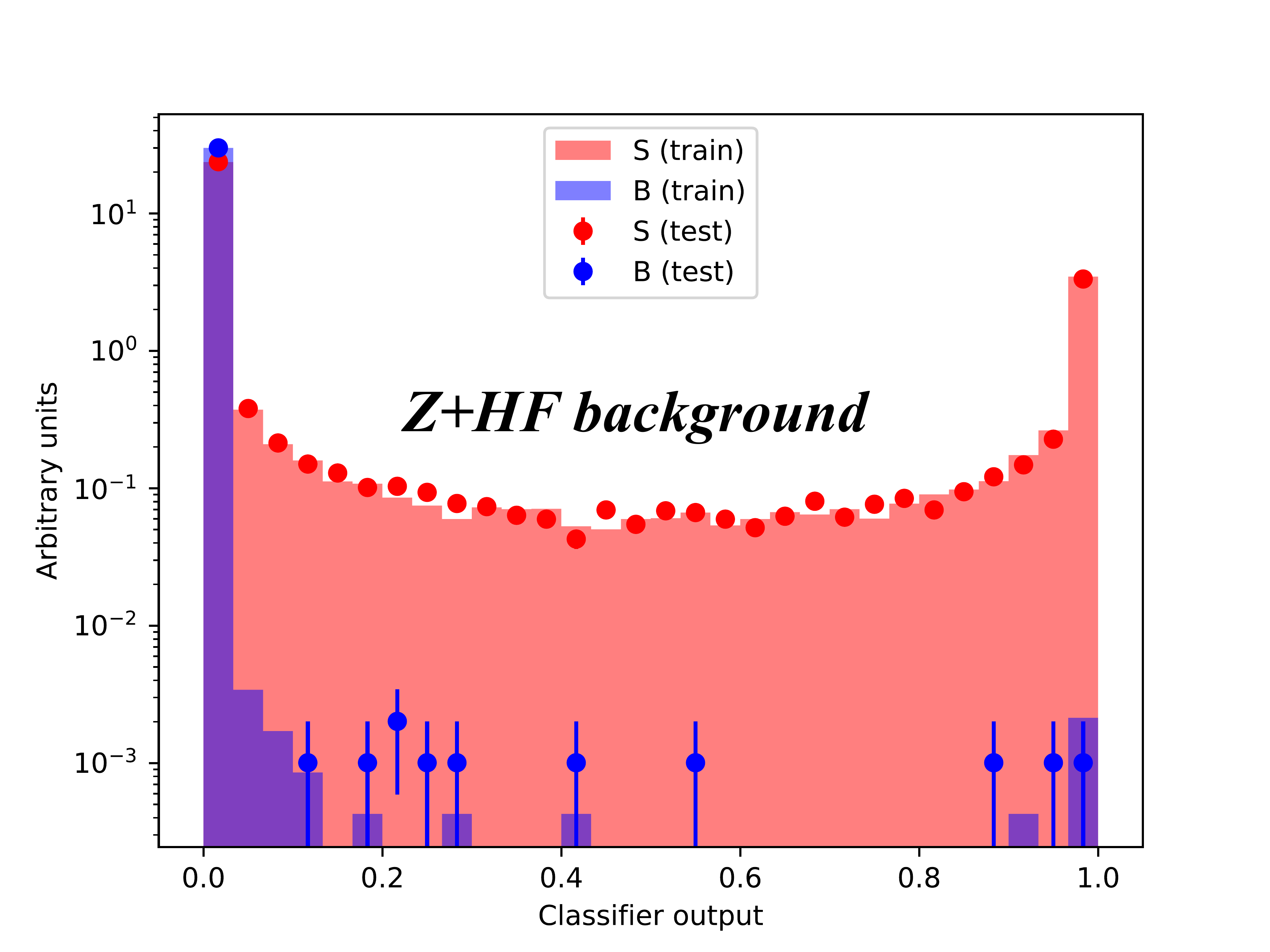}
    \end{center}
    \caption{Distribution of 0L signal (red) and background (blue) events as a function of  the classifier output. The left plot is the outcome of performing an initial pretraining run with 5 epochs. The middle (SM Higgs) and right (Z+HF) plots show the final distribution after a more precise, longer run is done. The solid distribution correspond to the outcomes on the training sample (70\% of the sample), whereas the dots correspond to the test sample. }\label{fig:testraining}
\end{figure*}

In Fig.~\ref{fig:testraining} we show the effect of pretraining in separating signal over background. The plots show the distribution of the  signal benchmark-point with $c_{HW} =0.03$ (red) and background (blue) events as a function of  the classifier output. The left plot is the outcome of performing an initial pretraining run with 5 epochs. The middle and right plots shows the final distribution after a longer run was performed. In all the plots, the solid distribution correspond to the outcomes on the training sample (70\% of the sample), whereas the dots correspond to the test sample. The fact that the solid distribution (train) and dots (test) distributions are similar is an indication that the algorithm is not overfitting. The middle plot compares BSM with SM Higgs production. The right plot is the separation between BSM and a reducible background, $Z$+HF, where a cut on the $m_{bb}$ variable in the Higgs mass window was done. By comparing the middle and right plot, one sees that the reducible background is easier to remove than the genuine SM Higgs background, as expected. 

\section{Results\label{results}}
\begin{figure*}[t!]%
    \begin{center}
        \includegraphics[scale = 0.23]{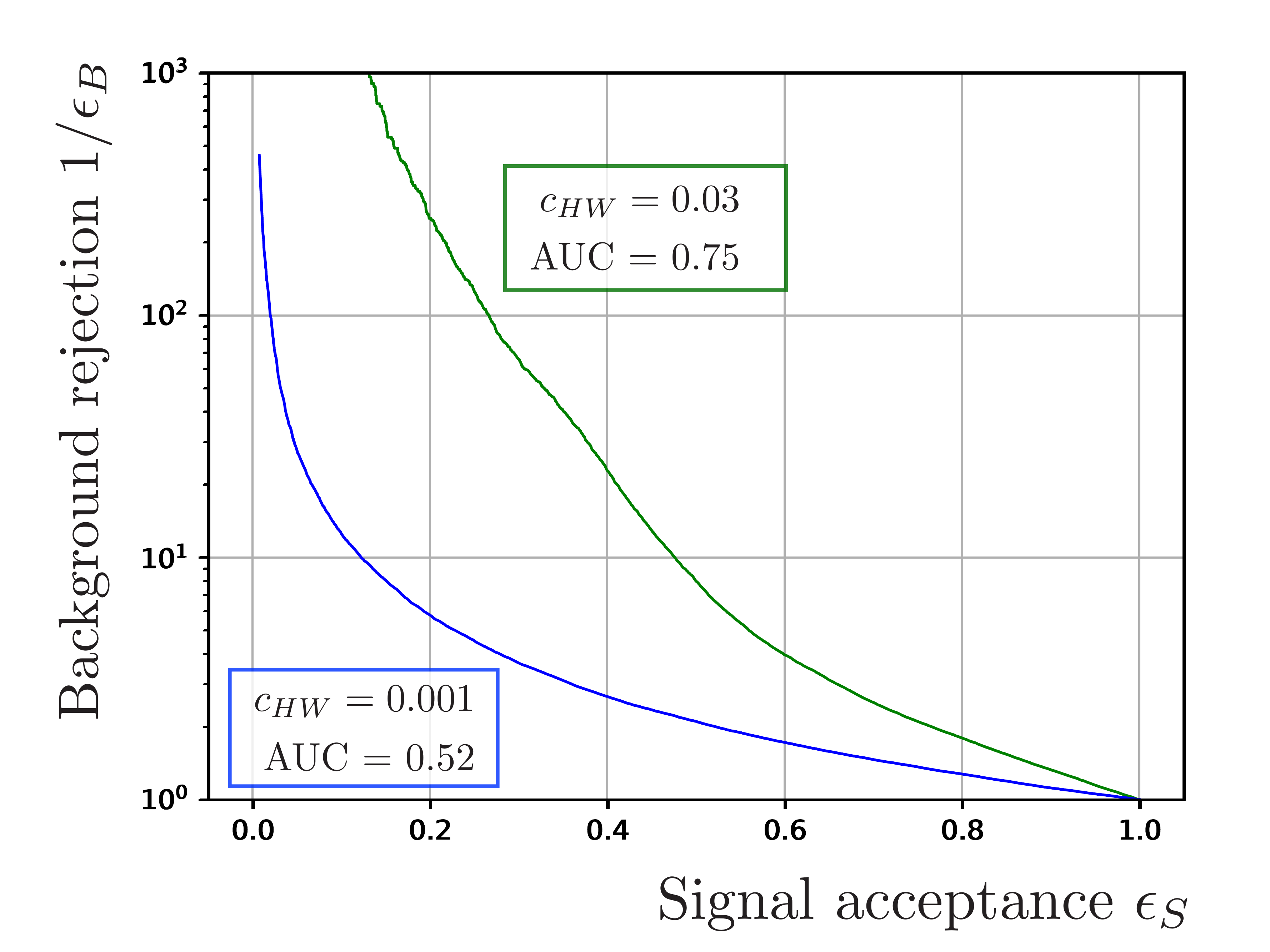}
  \includegraphics[scale = 0.23]{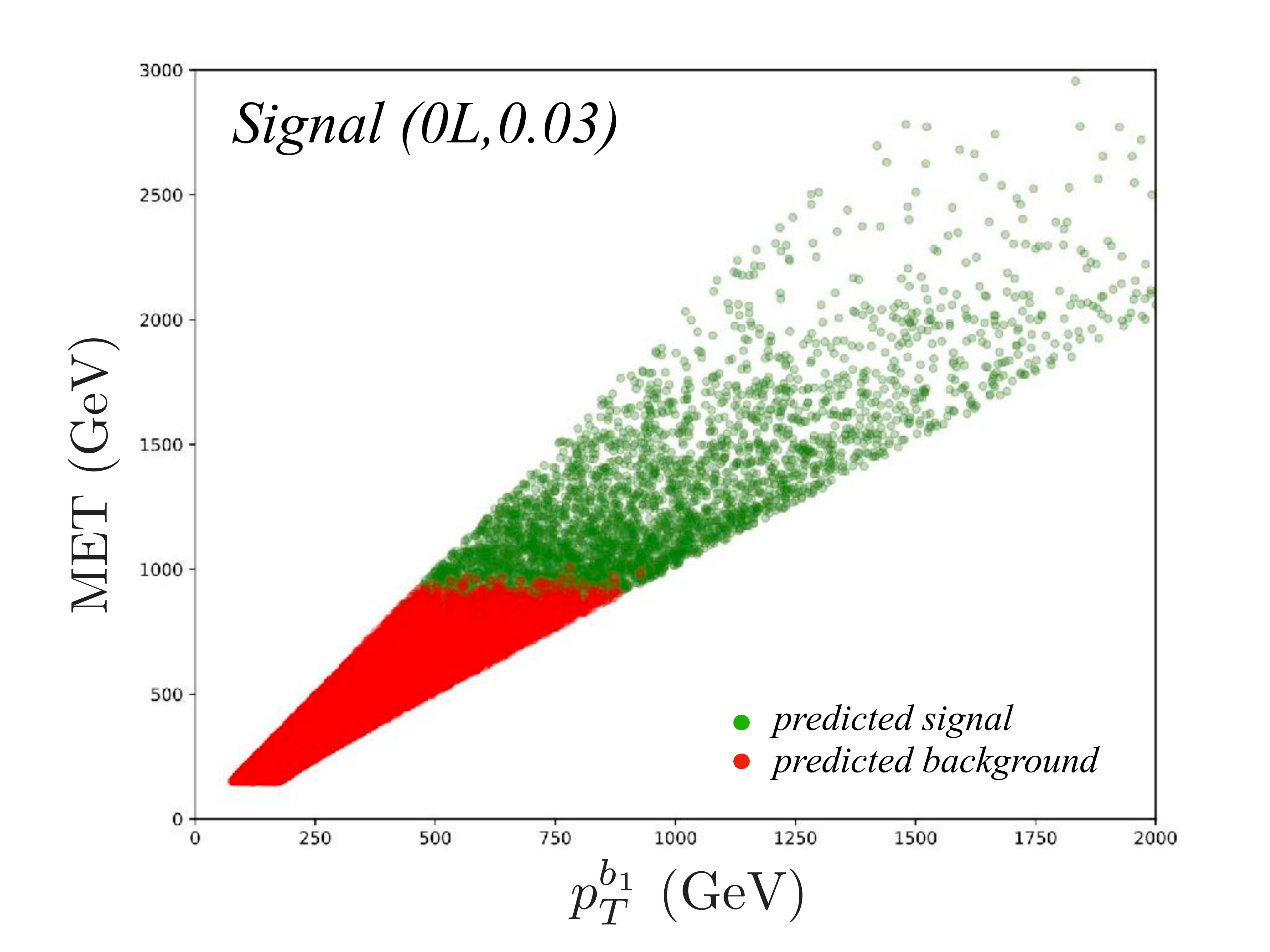}
     \end{center}
    \caption{Left: ROC curve for two values of the SMEFT coefficient in the 0L channel. Right: Classification of true signal events for $c_{HW}$=0.03 and their mapping to kinematic features.}\label{fig:roc}
\end{figure*}

The goodness of our procedure can be first evaluated by looking at the ROC curve in Fig.~\ref{fig:roc}, where we show the signal efficiency and background rejection curves. We present two examples of SMEFT effects, our benchmark value $c_{HW}$ = 0.03 and a very small value 0.001 which approaches the SM case. As expected, larger values of $c_{HW}$ present a better AUC and higher significance. 

Perhaps a more intuitive way to understand this ROC curve is to compute the predicted identity of events. In the right panel of Fig.~\ref{fig:roc}, we show a kinematic distribution of true signal events, separated by their predicted identity. Unsurprisingly, events with high energy are easier to distinguish from the SM backgrounds. This can be traced back to the Feynman rule in Eq.~\ref{eqcoupling}, where the SEMFT effects are momentum dependent and tend to lead to higher kinematic reach than SM interactions.

\begin{figure*}[t!]%
    \begin{center}
  \includegraphics[scale = 0.22]{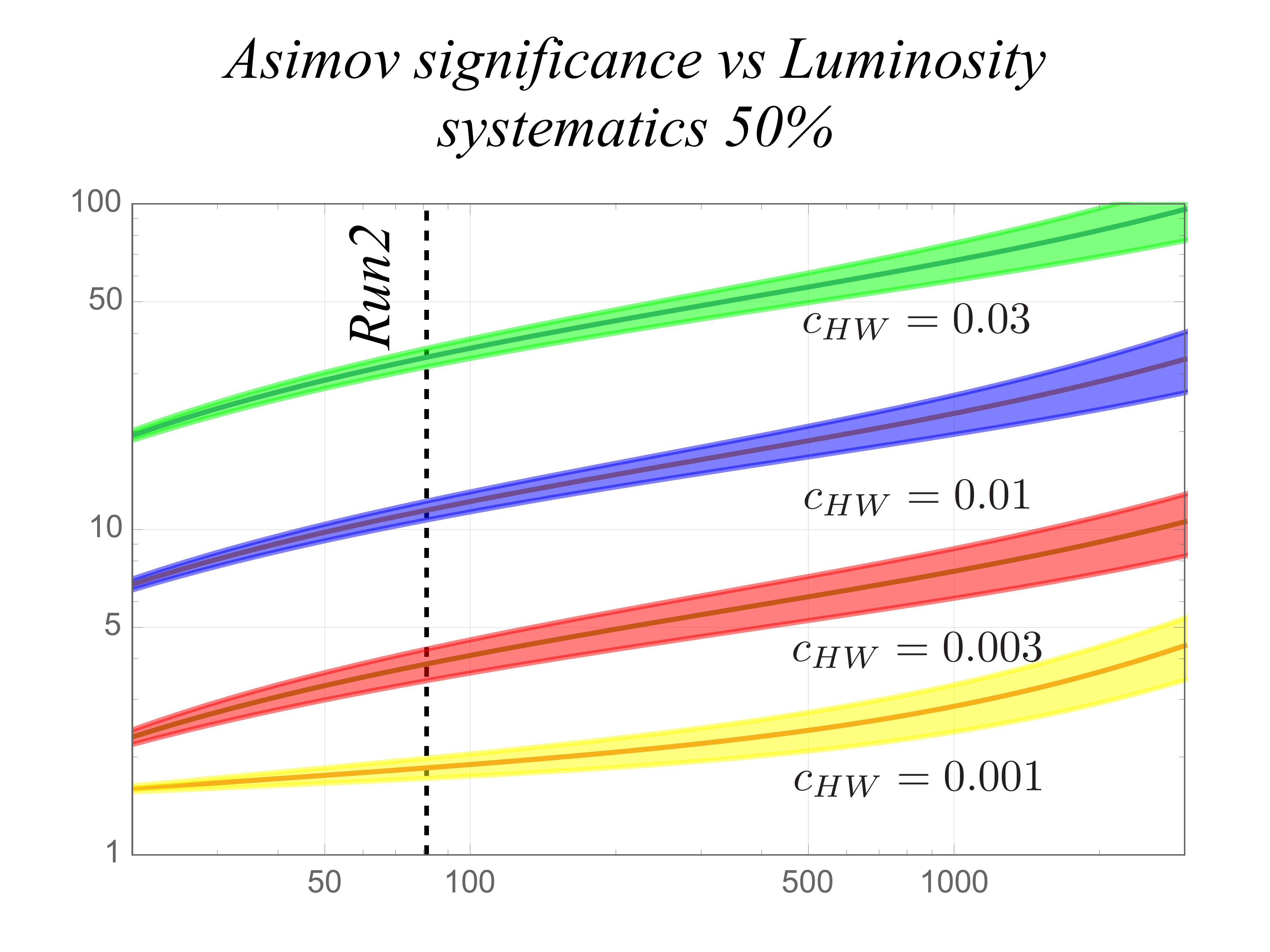}
   \includegraphics[scale = 0.22]{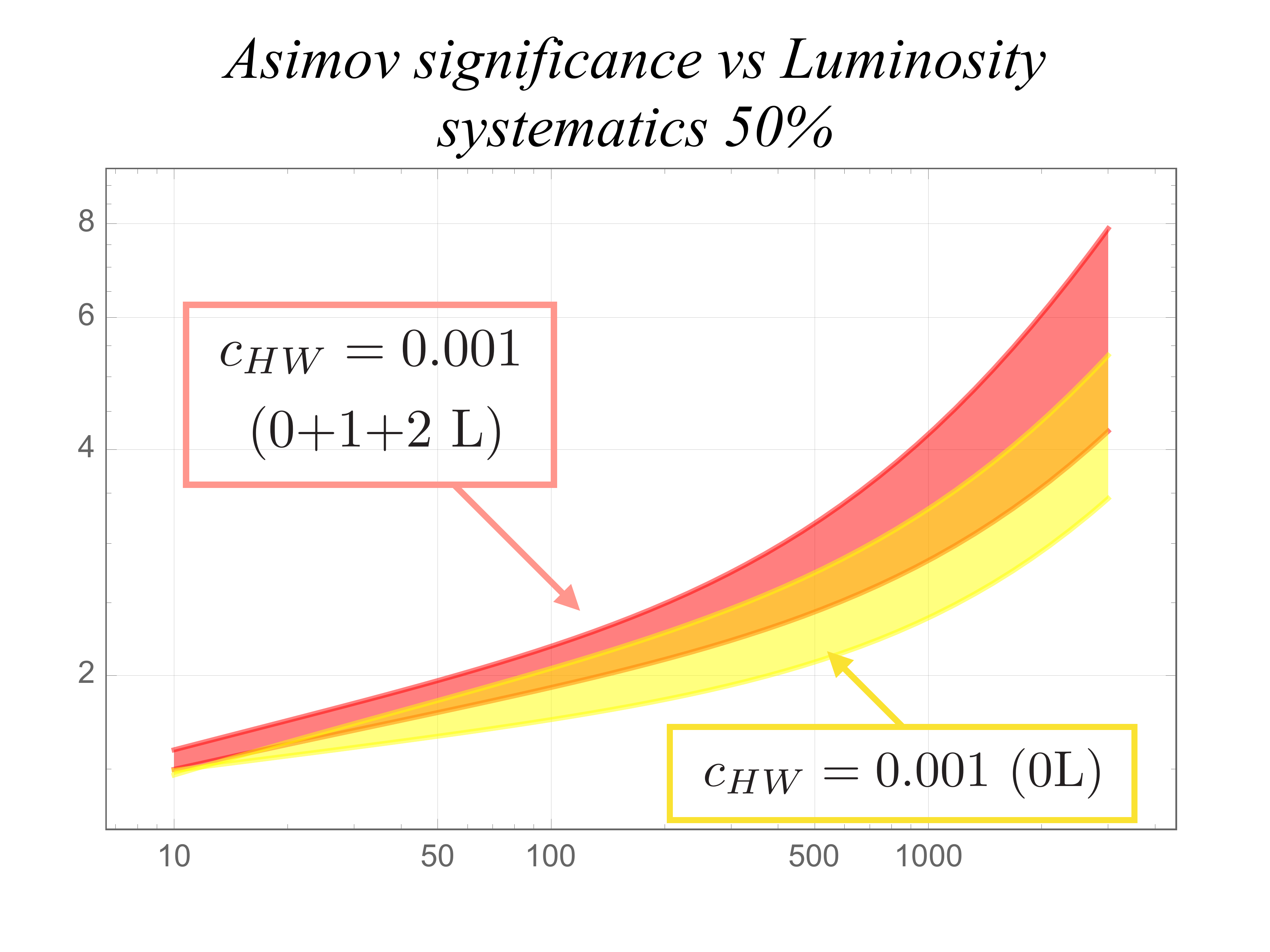}
    \end{center}
    \caption{Left: Luminosity (fb$^{-1}$) versus Asimov significance for different values of $c_{HW}$ in the 0-lepton channel and 50\% systematic uncertainty. Right: The effect of combination of VH channels for the limiting value $c_{HW}$=0.001 and 50\% systematic errors. }\label{fig:siglumi}
\end{figure*}

Nevertheless, quantities like the ROC curve and its AUC do not provide the answers we need in Particle Physics. We are interested in understanding beyond acceptance and rejection, but also the dependence with increasing luminosity and the effect of systematic uncertainties which are often disregarded in machine learning studies. In the right panel of Fig.~\ref{fig:siglumi} we show the Asimov significance in the 0L channel, for a choice of systematic uncertainty at 50\%, as a function of luminosity for various choices of the SMEFT coefficient. The bands correspond to 2$\sigma$ ranges. In contrast with the results from global fits, we obtain that values much below the 0.03 benchmark may be excluded by the Run2  data. The extent of this current exclusion cannot be obtained in a reliable fashion from our analysis, as we did perform an simplistic leading-order parton+shower analysis. Nevertheless, one would infer that sensitivity to value of $c_{HW}$ around 0.001 could be obtained using the  CMS and ATLAS combined Run2 data. 

 As one can see from the left panel of Fig.~\ref{fig:roc}, $c_{HW}$= 0.001 seems a limiting case for our algorithm in the 0L channel, as the AUC is very close to 0.5. We have chosen the 0L channel as it generally provides the best sensitivity to SMEFT~\cite{johnhiggscomplete}, but  one would wonder whether one could improve the sensitivity to this difficult point by combining with the other two channels (1L and 2L). The right panel in Fig.~\ref{fig:siglumi} shows the increase of sensitivity due to combination. For the small SMEFT deviation $c_{HW}=$0.001 and 50\% systematics, the improvement is within the error bars of the Asimov significance. A better handle on systematics could make the combination much more effective.
 
\section{Outlook}  \label{conclusions}
With the increasing experimental understanding of the LHC data, new ways to search for new physics open up. In particular, the use of detailed kinematic information is the next frontier in terms of LHC data characterisation. More capabilities come with more ambitions, particularly in terms of the complexity of new phenomena one can hope to tackle. We have identified one channel (VH) which is both relatively well understood and broad in terms of its kinematic reach, and a set of Machine Learning techniques which could allow us to detect new physics in the behaviour of the Higgs boson.  We chose the SMEFT as a template of the kind of deviations one could expect in the Higgs via virtual effects of new particles. 

Within the framework of our analysis, we found the 0L channel to be dominant, which was expected. We obtained a limit in the SMEFT coefficient $c_{HW} $ of 0.001, about 30 times better than the current constraint from a global analysis~\cite{johnnewest} with the Run2 data. This result shows the potential of incorporating these techniques to the SMEFT studies.

Our analysis could be improved in a number of ways. First and foremost, a more realistic simulation could be performed by the experiments, including NLO SMEF effects~\cite{NLOhel}. Secondly, although we found that deep layers led to overfitting, and a shallow NN was more suitable, new algorithms could be explored to increase sensitivity. In particular, one could use outlier detection without supervision. Thirdly, we should understand the effect of switching on more than one deviation along the lines described in Ref.~\cite{MLEFT}. This should be the stepping stone to a more global use of Machine Learning techniques in the area of global fits to SMEFT properties.


\begin{acknowledgments}
 F.F.F. is supported by project Y8Y2411B11 China Postdoctoral Science Foundation. C.K.K. wishes to acknowledge support from the Royal Society-SERB Newton International Fellowship (NF171488). The work of V.S. 
 by the Science Technology and Facilities Council (STFC) under grant number
ST/P000819/1.
\end{acknowledgments}

\appendix

\section{Glossary of terms}\label{glossary}
\begin{itemize}
\item[]   {\bf True positive rate ({\it tpr }$\equiv$ $\epsilon_S$) : }  ratio of true positive count and total signal events. The true positive counts are
 the number of signal events correctly identified by the algorithm. It also corresponds to the usual notion of signal acceptance.
\item[] {\bf False positive rate ({\it fpr }$\equiv$ $\epsilon_B$) :}  ratio of false positive counts and total number of background events.  The false positive counts are the background events, predicted as signal events by the algorithm. It also corresponds to the usual notion of  background rejection.
\item[] {\bf Receiver operating characteristic (ROC) : }  plot of {\it tpr} as a function of {\it fpr}   for each value of the classifier threshold between 0 to 1. 
\item[] {\bf Area under the curve (AUC) : }area under the ROC curve and a typical measure of the algorithm performance. 
\item[] {\bf Accuracy :} ratio of the correctly identified signal and background events versus total number of signal and background events.
\item[] {\bf Learning curve :} curve with shows the performance of the algorithm with iterative runs i.e. behaviour of the loss function with iterations. 
\item[] {\bf Batch :} data is divided into small sets, called batches, to save time and computation efforts.
\item[] {\bf Hidden Layers :} intermediate layers between  the input and output layers.
\item[] {\bf Asimov Significance :} Defined as a function of signal and background events and the uncertainty associated with the background ($\sigma_b$) 
\begin{eqnarray} \label{eq:asimov}
Z_{A} &=&\left[2\left((s+b)\ln\left[\frac{(s+b)(b+\bsvar)}{b^2+(s+b)\bsvar}\right] \right. \right. \nonumber -\\ &&
\left. \left. \frac{b^2}{\bsvar}\ln\left[1+\frac{\bsvar s}{b(b+\bsvar)}\right]\right)\right]^{1/2}.
\end{eqnarray}
\item[] {\bf Loss function :} the function which the algorithm searches to minimise. In our analysis, we used the 
following loss function designed to maximise the discovery significance : 
\begin{equation}
  \ell_{s/\sqrt{s+b}} = (s+b)/s^2,
\end{equation}
\begin{equation}
  \ell_{Asimov} = 1/Z_A
\end{equation}
\item[] {\bf Epochs :} The period between initialisation of the search for the minimum and when the batches pass  the NN. Basically, number of epochs is an 
iteration counter of how many times complete data set is explored by the algorithm, such that learning parameters are optimized.
\item[] {\bf Dropout: } mechanism to avoid the model overfitting, whereby the NN could drop few of the units (neurons) at the time of training.
\item[] {\bf Pretraining :} Quick pre-run with smaller number of epochs and steeper loss functions. The longer training is initialised by the pretraining hyperparameters.  
\item[] {\bf Classifier output :}  set of predictions for test sample. Our analysis is a binary classification problem, so with the predecided (user-decided) 
classification threshold, the events will either belong to signal or background class.
\end{itemize}

\section{Analysis set-up\label{setup}}
 We  generate 100K events for $WH$ and $ZH$ processes, with $\sqrt{s}$= 14 TeV, using MC$@$NLO madgraph6.3.2 and Higgs effective theory feynrules\cite{feynrules} model available in the literature \cite{benjalloul}. Note that in VH, the typical difference between 13 TeV and 14 TeV in terms of cross section is less than 10 \%.

We use Pythia6\cite{pythia6} to read the lhe events files but without doing the showering and hadronization. 
Considering leptonic decay channels of gauge boson and $b \bar b$ from Higgs decay, we have following three final states  
\begin{itemize}
\item 0-lepton (0L) :
\[ p p \rightarrow H Z, (H \rightarrow b \bar b, Z \rightarrow \nu \bar\nu)\]
\item 1 lepton (1L) :
\[ p p \rightarrow H W, (H \rightarrow b \bar b, W \rightarrow l v_l)\]
\item 2 lepton (2L) :
\[ p p \rightarrow H Z, (H \rightarrow b \bar b, Z \rightarrow l^+ l^-)\]

\end{itemize}
The events are  generated according to the ATLAS search strategy\cite{Hbbatlas} categorized as ``inclusive". The cuts applied are 
given in Table \ref{cuts} for all the channels.  We consider the two main
 background processes i.e. SM associated Higgs production with vector bosons and V $+$ heavy flavour (V$+$HF). 
 The SM associated Higgs production with vector bosons background 
is also generated importing the same model with $C_{HW}=0$.

We consider the following observables as data features:
\begin{itemize}
\item{$p_{T}^{b_{1}}$} transverse momentum of the leading b-jet.
\item{$p_{T}^{b_{2}}$} transverse momentum of the sub leading b-jet.
\item{$p_{T}^{VH}$} transverse momentum of the $VH$ pair.
\item{$M_{T}^{VH}$} transverse mass of the $VH$ pair.
\item{$p_{T}^{W/Z}$} transverse momentum of gauge boson.
\item{$p_{T}^H$} transverse momentum of the reconstructed Higgs boson.
\item{$\eta^H$} pseudo-rapidity of the reconstructed Higgs boson.
\item{$\phi^H$} azimuthal angle of the reconstructed Higgs boson.
\end{itemize}

which are common for all 3-channels and channel specific ones are :

\begin{itemize}
\item 0L channel : 
\begin{itemize}
\item $M_{T}^W$ transverse mass of the $W^{\pm}$ 
\item $p_{T}^l$ transverse momentum of lepton
\item $\slashed{E_{T}}$ missing transverse energy
\item $\Delta R_{wl}$ separation between lepton and $W$ boson in the $\eta-\phi$ plane
\item $\Delta \phi_{b_1l}$ azimuthal angular separation between leading b-jet and lepton 
\item $\Delta \phi_{l \slashed{E_T}}$ azimuthal angular separation between lepton  and $\slashed{E_T}$
\end{itemize}

\item 1L channel : 
\begin{itemize}
\item $\slashed{E_{T}}$ missing transverse energy
\item $\Delta \phi_{b_1 \slashed{E_T}}$ azimuthal angular separation between leading b-jet  and $\slashed{E_T}$
\end{itemize}

\item 2L channel : 
\begin{itemize}
\item $p_{T}^{l_1}$ transverse momentum of the leading lepton
\item $p_{T}^{l_2}$ transverse momentum of sub-leading lepton
\item $\Delta R_{ll}$ separation between two lepton in the $\eta-\phi$ plane
\item $\Delta \phi_{b_1 l_1}$ azimuthal angular separation between leading b-jet  and leading lepton
\item $\Delta \phi_{b_2 l_1}$ azimuthal angular separation between sub-leading b-jet  and leading lepton
\end{itemize}
\end{itemize}

\begin{table}[htb]
\centering
 $$
 \begin{array}{|c|c|}
\hline {\mbox{Channel}}  &{\mbox{Inclusive}}\\

\hline {\mbox{0L}} & { \slashed{E_T}>150\,\,\mbox{GeV}} \\
{\mbox{1L}} & {p_T^l>25 \,\,\mbox{GeV}, |\eta_l|<2.7 } \\
&  \slashed{E_T}>30\,\,\mbox{GeV}, p_T^V>150\,\,\mbox{GeV}  \\
{\mbox{2L}} & {p_T^l>7 \,\,\mbox{GeV},|\eta_l|<2.7 , p_T^V>75\,\,\mbox{GeV}}\\
& \mbox{Leading lepton $p_T>$ 27 GeV}  \\
  \hline {\mbox{0L,1L,2L}} & p_T^b>20 \,\,\mbox{GeV}, |\eta_b|<2.5 , \\
 & {\mbox{Leading b-jet $p_T>$ 45 GeV} } \\
\hline 
\end{array}
 $$
\caption{Cuts applied at event generation level for both signal and background process. In case 
of Z+HF we apply an additional cut on $m_{b\bar b}$ i.e. 115 $< m_{b \bar b}<$ 135 GeV.}
 \label{cuts} \end{table}


\end{document}